\documentclass[prb,aps,epsfig,showpacs]{revtex4}
\usepackage{amsmath} 
\usepackage{amsfonts} 
\usepackage[dvips]{graphicx,color} 
\usepackage{subfigure}  
\begin{document}
\sloppy
\title
{Collapse dynamics of a ${}^{176}\textrm{Yb}\,$-${}^{174}\textrm{Yb}$
Bose-Einstein condensate
}
\author{G. K. Chaudhary}  
\email{gkchaudhury@physics.du.ac.in} 
\author{R. Ramakumar}
\email{rkumar@physics.du.ac.in} 
\affiliation
{Department of Physics and Astrophysics, University of Delhi,
Delhi-110007, India} 
\date{4 June 2010}
\begin{abstract}
In this paper, we present a theoretical study of a two-component Bose-Einstein 
condensate composed of Ytterbium (Yb) isotopes in a three dimensional
anisotropic harmonic potential. The condensate consists of a 
mixture of ${}^{176}\textrm{Yb}$ atoms which have a negative s-wave
scattering length and ${}^{174}\textrm{Yb}$ atoms having a positive
s-wave scattering length. We study the ground state as well as dynamic 
properties of this two-component condensate. Due to the attractive 
interactions between ${}^{176}\textrm{Yb}$ atoms, the 
condensate of ${}^{176}\textrm{Yb}$ undergo
a collapse when the particle number exceed a critical value. 
The critical number and the collapse 
dynamics are modified due to the presence of ${}^{174}\textrm{Yb}$ atoms.
We use coupled two-component Gross-Pitaevskii equations to study the collapse
dynamics. The theoretical results obtained are in reasonable
agreement with the
experimental results of 
Fukuhara {\em et al.} [PRA{\bf 79}, 021601(R) (2009)].
\end{abstract}
\maketitle
\narrowtext
\section{Introduction}
\label{sec1}
The first experimental observation of Bose-Einstein condensate (BEC) 
\cite{anderson,davis,bradley}
in bose atom vapors have initiated an exciting field of research, 
both theoretically and experimentally. One of the most 
interesting developments in this field is the formation of multi-component 
condensates. Multi-component BECs have been observed experimentally 
by Myatt {\em et al.} \cite{myatt} and Hall {\em et al.}\cite{hall} 
for two different hyperfine spin sates of ${}^{87}\textrm{Rb}$,
by Modugno {\em et al.} \cite{modungo} for different atoms 
$({}^{41}\textrm{K}$ and ${}^{87}\textrm{Rb})$, and 
Papp {\em et al.} \cite{papp} for different isotopes of the same atom  
$({}^{85}\textrm{Rb}$ and ${}^{87}\textrm{Rb})$. A rich variety of various 
interesting effects exhibited by these two-component BECs 
have inspired a number 
of theoretical studies covering various aspects of a these systems 
\cite{ho,pu,goral,adhikari,kasamatsu2,tsubota}. The common feature of
the bose systems in 
these experiments is that the intra-component and the inter-component
boson-boson interactions are all repulsive. It rises 
the curiosity about the properties of a multi-component condensate 
in which one kind of atoms have  
repulsive interactions while another kind of atoms have
attractive interactions. Recently, Fukuhara {\em et al.}\cite{fukuhara} 
observed BEC of spin-zero $\textrm{Yb}$ isotopes by
implementing an all-optical cooling protocol. 
The bose-bose mixture in these experiments contain ${}^{174}\textrm{Yb}$ 
atoms having a positive s-wave scattering length and ${}^{176}\textrm{Yb}$ 
atoms having a negative s-wave scattering length\cite{fukuhara,kitagawa}. 
The s-wave scattering length between ${}^{174}\textrm{Yb}$  
and  ${}^{176}\textrm{Yb}$ is also positive\cite{kitagawa}. 
Such a two-component condensate can be expected to show dynamical
properties far more complex than a one-component condensate
of attractively interacting bosons, which becomes unstable
when the number of atoms exceed a critical value\cite{bradley,holland,donley}.
\par
For a two-component BEC with attractive interactions between
bosons in one component and repulsive interactions in the second
component,
two most basic questions are of that of its stability and that of the
collapse dynamics.
In this paper, we theoretically study the ground 
state as well as dynamic properties of such a two-component 
condensate in a three dimensional (3D) anisotropic harmonic potential.
We specifically consider the ${}^{174}\textrm{Yb}$-${}^{176}\textrm{Yb}$
bose-bose mixture in the anisotropic harmonic confining potential
as in the experiment of Fukuhara {\em et al.}\cite{fukuhara}.
The paper is arranged as follows. In Sec. II, we describe the theoretical model
for the study of a two-component BEC.  In Sec. III, we discuss the ground state
profile of a stable two-component BEC. In Sec. IV, we present
the collapse dynamics of the system and compare it with
the experimental results\cite{fukuhara}.
The conclusions are given in Sec. V.
\section{Two-Component BEC: Theoretical model}
\label{sec2}
The ground state and dynamic properties of a two component Bose-Einstein 
condensate (BEC)is well described by a set of coupled
Gross-Pitaeveskii equations \cite{gross,pitaevskii,sinatra} (GPE) given by 
\begin{equation}
 i \hbar\frac{\partial\psi_{i}({\bf r},t)}{\partial t}=(-\frac{\hbar^2}{2m_{i}}
\nabla^2+v_{i}(r)
+ g_{ii}\vert\psi_{i}({\bf r},t)\vert^2
+g_{ij}\vert\psi_{j}({\bf r},t)\vert^2)\psi_{i}({\bf r},t)\,,
\label{NLS_eqn}
\end{equation}
where $i=1,2$ are indices for the two components $($$1$ for ${}^{174}\textrm{Yb}$ 
and $2$ for ${}^{176}\textrm{Yb}$$)$, $j=3-i$, ${\bf r} \equiv (x,y,z)^T$ 
is the spatial coordinate vector, $v_{i}(r)=(1/2)m_{i}(\omega_{x}^2x^2+
\omega_{y}^2y^2+\omega_{z}^2z^2$) 
is the trapping potential, $g_{ii}=4\pi\hbar^2a_{ii}/m_{i}$ is the intra-species 
interaction and $g_{ij}=2\pi\hbar^2a_{ij}/m_{ij}$ the inter-species 
interaction strength between atoms in the condensed state, $a_{ii}$ is 
intra-species and $a_{ij}$ is inter-species s-wave scattering length, 
$m_{ij}=m_{i}m_{j}/(m_{i}+m_{j})$ is the reduced mass in which
$m_i$ and $m_j$ are atomic masses. 
The normalization condition for each component is  
$\int\vert\psi_{i}(r)\vert^2dr=N_{i}\,,$
where $N_{i}$ is number of atoms in each component.
\par
We non-dimensionalize 
Eq.$\,$(\ref{NLS_eqn}) through a set of linear 
transformations: $\tilde{t}=\omega_{x} t,\:\:\tilde{{\bf r}}={\bf r}/l,\:\:
\widetilde{\psi_{i}}({\bf r})=N_{i}^{-\frac{1}{2}}l^{3/2}\psi({\bf r})$. After 
dropping the wiggles on the symbols, 
we obtain
\begin{equation}
 i \frac{\partial\psi_{i}({\bf r},t)}{\partial t}=(-\frac{1}{2}\nabla^2+
v_{i}({\bf r})
+ \lambda_{ii}\vert\psi_{i}({\bf r},t)\vert^2
+\lambda_{ij}\vert\psi_{j}({\bf r},t)\vert^2)\psi_{i}({\bf r},t)\,,
\label{NLS_non_dim_eqn}
\end {equation}
where
\begin{equation}
l=\sqrt{\frac{\hbar}{m \omega_{x}}}, \:\:\lambda_{ii} = \frac{4\pi a_{ii}}{l},
\:\:\lambda_{ij} = \frac{4\pi a_{ij}}{l}\,,\nonumber \\
\label{constants}
\end{equation} 
\begin{equation}
v_{i}({\bf r})=\frac{1}{2}(x^2+\kappa^2y^2+\gamma^2z^2), 
\:\:\kappa=\frac{\omega_{y}}{\omega_{x}},\:\:\gamma=\frac{\omega_{z}}{\omega_{x}}.
\end{equation} 
Since the masses of the two isotopes are nearly equal, we have taken
$m_1=m_2=m$.
In order to find a stationary solution of 
Eq.$\,$(\ref{NLS_non_dim_eqn}), we do a separation of variables
$\psi_{i}({\bf r},t)=\psi_{i}({\bf r}) \times{\text {exp}}[-i (\mu_{i}/(\hbar
\omega_{x}))t]$,
where $\mu_{i}$ is the chemical potential of the $i$th component. Starting 
from Eq.$\,$(\ref{NLS_non_dim_eqn}), we obtain
\begin{equation}
(-\frac{1}{2}\nabla^2+v_{i}({\bf r})+ \lambda_{ii}\vert\psi_{i}({\bf r})
\vert^2+\lambda_{ij}\vert\psi_{j}({\bf r})\vert^2)\psi_{i}({\bf r}) 
=\frac{\mu_{i}}{m \omega_{x}}\psi_{i}({\bf r})\,.
\label{stationary_state_eqn}
\end {equation}  
\section{Ground state profiles of a two-component BEC of Yb atoms}
\label{sec3}
\begin{figure}
\resizebox*{3.5in}{3.0in}{\rotatebox{270}{\includegraphics{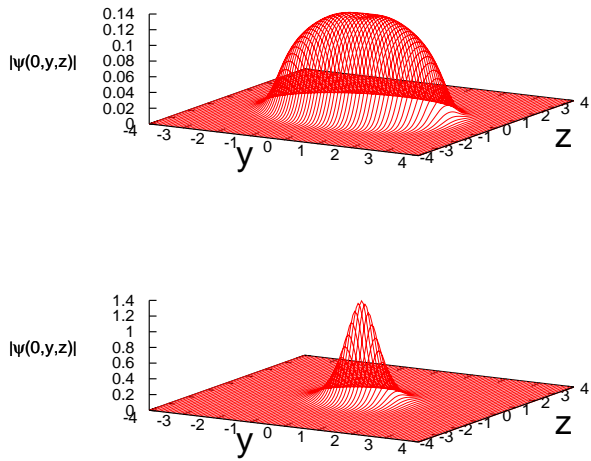}}}
\vspace*{0.5cm}
\caption[]
{Ground state profiles of ${}^{174}\textrm{Yb}$ (top panel) and 
${}^{176}\textrm{Yb}$ (bottom panel) condensates in a 3d anisotropic harmonic
potential with $N_{1}=6\times 10^4$ and $N_{2}=150$.}
\label{scaling}
\end{figure}
In this section, we discuss the ground state properties of a two-component BEC of 
Yb isotopes by numerically solving the coupled GPE (Eq.$\,$\ref{stationary_state_eqn}). 
The ground state solution of the GPE is found  by the imaginary time 
propagation method. In this method, the time dependent GPE 
is evolved in imaginary time starting from an initial guess 
using a finite difference Crank-Nicholson (FDCN) scheme\cite{
muruganandan}. In imaginary time propagation we have 
taken the space step as $\delta x=\delta y=\delta z=0.1$ and 
the time step as $\delta t=0.00005$.
We have used a set of parameters corresponding to 
the ${}^{174}\textrm{Yb}$\,-${}^{176}\textrm{Yb}$
system in the experiment\cite{fukuhara}: $m\,=\, 2.8734238\times10^{-25}\,Kg$, $a_{11}\,=\,5.55\times10^{-9}\, 
m$, $a_{22}\,=\,-1.28\times10^{-9}\, m$, $a_{12}\,=\,a_{21}\,=\,2.88\times10^
{-9}\, m$, $\nu_{x}\,(=\omega_{x}/2\pi)\,=\,45\,Hz$,
$\nu_{y}\,(=\omega_{y}/2\pi)\,=\,200\,Hz$, $\nu_{z}\,(=\omega_{z}/2\pi)\,=\,
300\,Hz$.
\par
Due to the attractive interaction between ${}^{176}\textrm{Yb}$ atoms, the condensate 
of ${}^{176}\textrm{Yb}$ undergo a collapse if the particle number exceeds a critical value 
$N_{2c}$ \cite{holland,donley}. For a single component 
${}^{176}\textrm{Yb}$ this value is given 
by $N_{2c}\approx 0.5L/{\vert a \vert}$, where
$L=\sqrt{\hbar/m \omega}$ and 
$\omega=(\omega_{x}\omega_{y}\omega{z})^{\frac{1}{3}}$. 
For the parameters given above, $N_{2c}\approx 250$.
But, the critical number is modified due to the 
presence of ${}^{174}\textrm{Yb}$ 
atoms with a positive scattering length.
The physical origin of this modification is in the effective
potential felt by the bosons in the attractive component.
The interaction with the repulsive component changes the
effective potential from 
$v_{2}({\bf r})-|\lambda_{22}|\vert\psi_{2}({\bf r})\vert^2$
to $v_{2}({\bf r})-|\lambda_{22}|\vert\psi_{2}({\bf r})
\vert^2+\lambda_{21}\vert\psi_{1}({\bf r})\vert^2$.
This reduces the effect of the attractive interaction and leads
to a reduction of the critical number.
Alternatively, one may say that the mean-field contribution from the repulsive 
component leads to a flattening of the effective potential
felt by the attractive component.
If $N_{1}=6\times 10^4$, then for the given parameters,
$N_{2c}$ is calculated to be $\simeq220$ by our numerical simulation. 
In order to prepare a stable condensate, we have 
taken $N_{1}=6\times 10^4$ and $N_{2}=150$.
The ground sate profile of the two-component BEC is presented 
in Fig. 1. We observe that the ${}^{176}\textrm{Yb}$ atoms 
are at the center of trap
and are surrounded by a large cloud of ${}^{174}\textrm{Yb}$ atoms.
\section{Collapse Dynamics of a two-component BEC of Yb atoms}
\label{sec3}
\begin{figure}
\resizebox*{3.5in}{2.50in}{\rotatebox{270}{\includegraphics{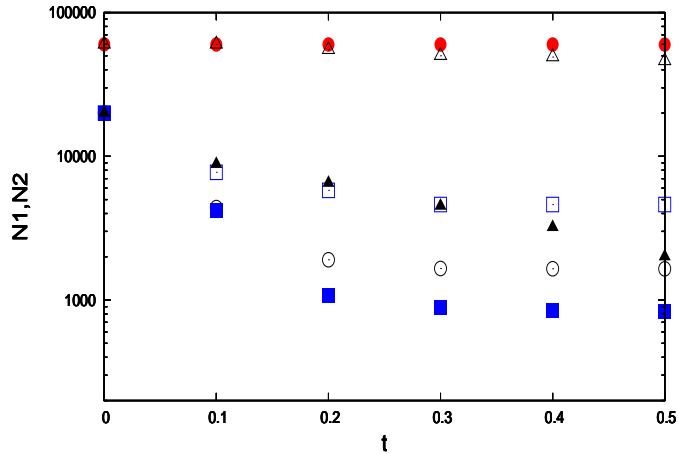}}}
\vspace*{0.5cm}
\caption[]
{Time evolution of number of ${}^{174}\textrm{Yb}$(filled circles), 
${}^{176}\textrm{Yb}$ (open squares, open circles, filled squares). 
The corresponding experimental results are given 
by (open triangles for ${}^{174}\textrm{Yb}$ and 
filled triangles for ${}^{176}\textrm{Yb}$). The values
of three body recombination terms are:
$K_{3}^1=4.2\times 10^{-29}{cm}^6 s^{-1}$(for all points)), 
$K_{3}^2=3.0\times 10^{-28}{cm}^6 s^{-1}$(open squares),
$K_{3}^2=3.0\times 10^{-27}{cm}^6 s^{-1}$(open circles),
$K_{3}^2=3.0\times 10^{-26}{cm}^6 s^{-1}$(filled squares). The time is in 
units of seconds.}
\label{scaling}
\end{figure}
In this section, we study the collapse dynamics of a two component BEC
composed of ${}^{176}\textrm{Yb}$ and ${}^{174}\textrm{Yb}$ atoms
using the coupled time-dependent GPE's.  Due to the 
negative  s-wave scattering length, the ${}^{176}\textrm{Yb}$ 
condensate becomes unstable if  the number of atoms becomes 
greater than a critical value $N_{2c}$. This condensate collapses
by emitting atoms out of it. 
Due to high density of atoms in the attractive condensate, the loss 
of atoms from
the condensate occurs through three-body collisions.
To model  this collapse we add a imaginary three body quintic loss term   
\cite{kagan1,kagan2,saito,duine} to 
the RHS of GPE Eq.$\,$(\ref{NLS_eqn}) given by  
\begin{equation}
K_{i}\psi_{i}=-(1/12)i K_{3}^i \vert \psi_{i} \vert^4 \psi_{i}\,,
\end{equation}
where $K_{3}^i$ is the three-body loss coefficient for each component.
We have neglected the two-body dipolar loss term  as 
they make negligible contribution in this 
case\cite{fukuhara,kagan2,saito,duine}. We have also left out the loss 
terms for ${}^{174}\textrm{Yb}-^{174}\textrm{Yb}-^{176}\textrm{Yb}$, 
$^{174}\textrm{Yb}-^{176}\textrm{Yb}-^{176}\textrm{Yb}$,
$^{176}\textrm{Yb}-^{176}\textrm{Yb}-^{174}\textrm{Yb}$,
and $^{176}\textrm{Yb}-^{174}\textrm{Yb}-^{174}\textrm{Yb}$ collisions,
since the losses due to these are comparatively small \cite{fukuhara}.
So, Eq.$\,$(\ref{NLS_non_dim_eqn}) becomes 
\begin{equation}
 i \frac{\partial\psi_{i}({\bf r},t)}{\partial t}=(-\frac{1}{2}\nabla^2+
v_{i}({\bf r})
+ \lambda_{ii}\vert\psi_{i}({\bf r},t)\vert^2
+\lambda_{ij}\vert\psi_{j}({\bf r},t)\vert^2-i \xi_{i}\vert\psi_{i}({\bf r},t)\vert^4)\psi_{i}({\bf r},t),
\label{quintic gpe}
\end {equation}
where
$\xi_{i}=(1/12) N^2 K_{3}^i l^{-6} \omega_{x}^{-1}.$
\par
We time evolve the coupled time-dependent GP equations 
Eq.$\,$(\ref{quintic gpe}) using  the finite difference Crank-Nicholson (FDCN) scheme 
with a known initial condition. In real time propagation,
we have taken the space step as $\delta x=\delta y=\delta z=0.1$ 
and the time step as $\delta t=0.005$.
The time evolution of the number of
${}^{176}\textrm{Yb}$ and  ${}^{174}\textrm{Yb}$
is shown in Fig. 2 along with the experimental 
data of Fukuhara {\em et al.}\cite{fukuhara}. Considering the complex
dynamics of the the two-component system with mixed interactions, the 
theoretical results may be said to be in reasonable 
agreement with the experimental results.
We observe that there is a significant loss of 
${}^{176}\textrm{Yb}$ atoms. We also see that the decay of 
${}^{176}\textrm{Yb}$ is very rapid. 
It is due to the collapse of the  ${}^{176}\textrm{Yb}$ condensate. 
The number of ${}^{174}\textrm{Yb}$ atoms does show a very small
decrease, which is not visible on the scale of this figure.
To understand the details of the decay process, we study 
the condensate profiles of each 
component at different times. 
The results are shown in the Fig. 3 for  ${}^{176}\textrm{Yb}$ 
and in Fig. 4 for ${}^{174}\textrm{Yb}$.
At $t=0$, the ${}^{176}\textrm{Yb}$ are at
the center of the trap (top left panel of Fig. 3) 
surrounded by ${}^{174}\textrm{Yb}$ (top left panel of Fig. 4).
Since the number of atoms in the attractive component is higher 
than the critical number for stability, the system is unstable.
When the system evolves in time, the attractive component
explodes as is evident from the spreading of this
component with time, in real space, as shown in Fig. 3. 
The spiky structures in these figures
represent the inhomogeneities produced due to the on-going
explosion process. As mentioned earlier, the explosion
also leads to a spread of the ground state profile.
Due to the coupling between the
attractive and repulsive components, the condensate
of ${}^{174}\textrm{Yb}$ is also redistributed in real
space during the time evolution, as shown in Fig. 4. 
The numbers of remaining atoms in each condensate component 
during the time evolution, for a longer period of time,
is shown in Fig. 5.
We note that the best agreement with the experimental
results (Fig. 2) is obtained for the measured values of the $K^{1}_{3}$
and $K^{2}_{3}$. The disagreement at later times is likely to be
originating from the neglect of atomic loss due to collisions involving
${}^{176}\textrm{Yb}$ and ${}^{174}\textrm{Yb}$. During
the initial stages of the time evolution, the bosons distribution
gets heavily mixed due to the explosion and due to the coupling
between the two components. Then, at later times, the inter-component
collisions is likely to affect the atom loss. We are
unable to include these loss terms since their values are not known
at present. 
\begin{figure}
\resizebox*{3.5in}{3.0in}{\rotatebox{270}{\includegraphics{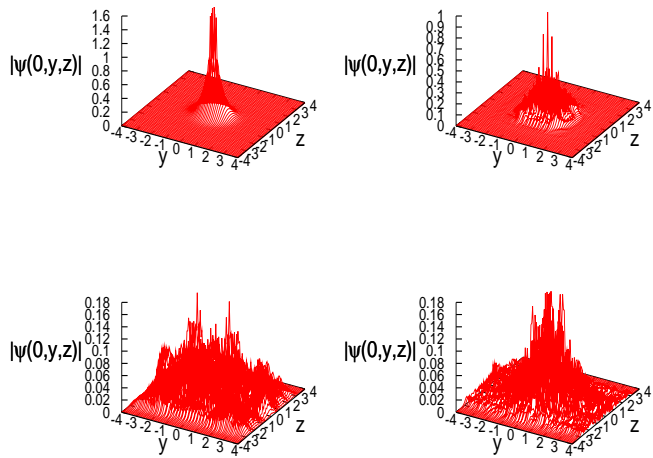}}}
\vspace*{0.5cm}
\caption[]
{Ground state profile of ${}^{176}\textrm{Yb}$ at different times.
The four figures correspond to $t (seconds) = 0.0$ (top left panel), 
$0.2$ (top right) $0.4$ (bottom left), and $2$ (bottom right). The values of 
three body recombination terms are: $K_{3}^1=4.2\times 10^{-29}{cm}^6 s^{-1}$,
$K_{3}^2=3.0\times 10^{-28}{cm}^6 s^{-1}$. Here,
$N_{1}=6\times 10^4$ and $N_{2}=2\times 10^4$.}
\label{scaling}
\end{figure}
\begin{figure}
\resizebox*{3.5in}{3.0in}{\rotatebox{270}{\includegraphics{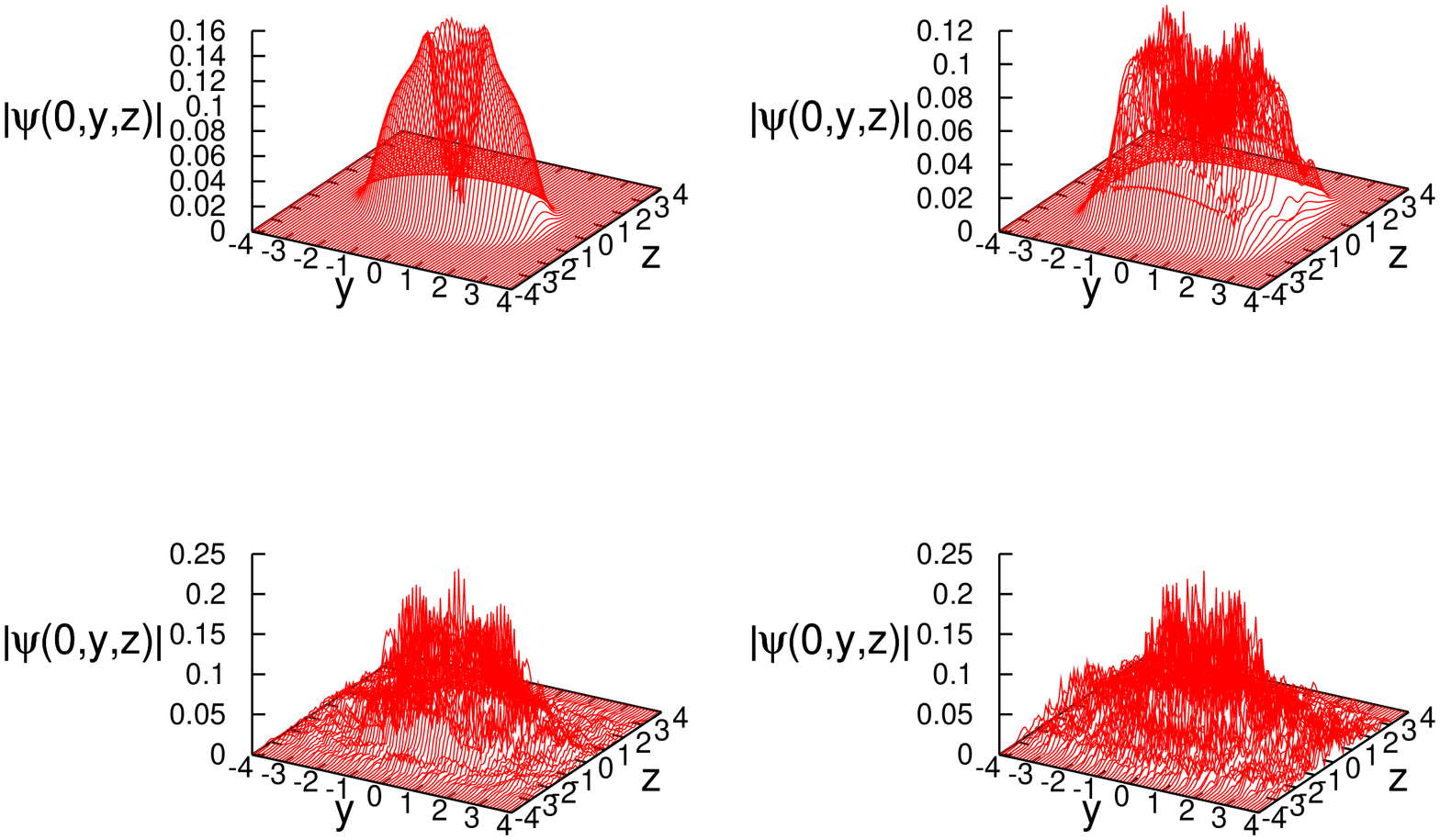}}}
\vspace*{0.5cm}
\caption[]
{Ground state profile of ${}^{174}\textrm{Yb}$ at different times.
The four figures correspond to $t (seconds) = 0.0$ (top left panel), 
$0.2$ (top right) $0.4$ (bottom left), and $2$ (bottom right). 
The values of the three body recombination terms 
are: $K_{3}^1=4.2\times 10^{-29}{cm}^6 s^{-1}$,
$K_{3}^2=3.0\times 10^{-28}{cm}^6 s^{-1}$. Here,
$N_{1}=6\times 10^4$ and $N_{2}=2\times 10^4$.}
\label{scaling}
\end{figure}
\begin{figure}
\resizebox*{3.5in}{3.0in}{\rotatebox{270}{\includegraphics{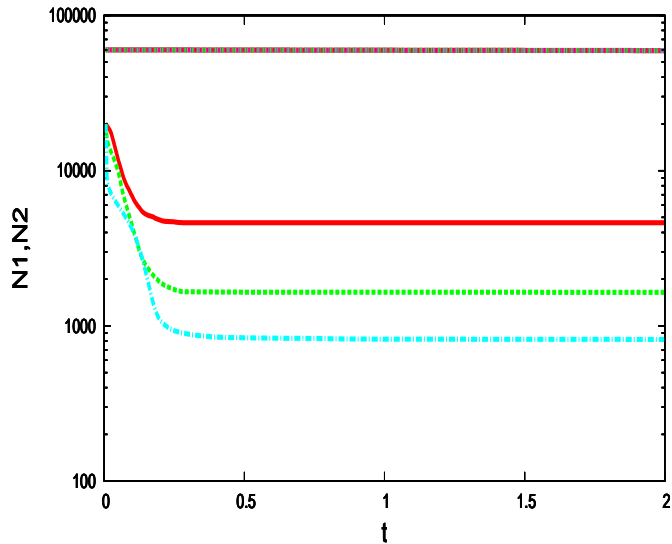}}}
\vspace*{0.5cm}
\caption[]
{Time evolution of number of ${}^{174}\textrm{Yb}$(top curve), 
${}^{176}\textrm{Yb}$(solid line, dashed, dash-dot lines). 
The values of three body recombination terms are: 
$K_{3}^1=4.2\times 10^{-29}{cm}^6 s^{-1}$ (for all lines)), 
$K_{3}^2=3.0\times 10^{-28}{cm}^6 s^{-1}$ (solid line),
$K_{3}^2=3.0\times 10^{-27}{cm}^6 s^{-1}$ (dashed line),
$K_{3}^2=3.0\times 10^{-26}{cm}^6 s^{-1}$ (dash-dot line). 
The time is in units of seconds.}
\label{scaling}
\end{figure}
\section{conclusions}
\label{sec5}
In this paper, we presented a study of some static and dynamic properties
of a two-component bose condensate
consisting of repulsively interacting ${}^{174}\textrm{Yb}$ atoms
and attractively interacting ${}^{176}\textrm{Yb}$ atoms in an
anisotropic harmonic confinement.
In the stable state, the ground state has ${}^{176}\textrm{Yb}$
atoms in the center of the trap surrounded by ${}^{176}\textrm{Yb}$ atoms.
When the number of atoms in the attractive component exceed a critical
value, the the system undergo a collapse. We analyzed the time
evolution of this collapse process for the specific system
parameters of the ${}^{176}\textrm{Yb}\,$-${}^{174}\textrm{Yb}$
system studied in the experiment of Fukuhara {\em et al.}.
The details of the collapse dynamics
are found to be in reasonable agreement with experimental results.
The critical number for stability of the attractive condensate is
reduced by it's interaction with the repulsive condensate.
\section{Acknowledgments}
\label{sec6}
Gopesh Kumar Chaudhary thanks UGC, Government of India, for  
financial support during this work. 
GKC thanks Prof. G. V. Shlyapnikov
for a discussion during the ICTS International School on Cold Atoms 
and Ions held recently in Kolkata.  We thank Prof. J. K. Bhattacharjee
for a discussion during an SERC school on Nonlinear Dynamics held
recently in Delhi.

\end{document}